\documentclass[runningheads]{llncs}
\usepackage[T1]{fontenc}

\usepackage{graphicx}
\usepackage{amsmath}
\usepackage{amsfonts}
\usepackage{booktabs}
\usepackage{makecell}
\usepackage{multirow}
\usepackage{color}
\usepackage{verbatim}
\usepackage[misc]{ifsym}
\newcommand{\littleshrinker}{\vspace{-0.25cm}}

\newcommand{\subtleshrinker}{\vspace{-0.04cm}}

\makeatletter
\def\blfootnote{\xdef\@thefnmark{}\@footnotetext}
\makeatother

\begin{document}
\title{Improving Micro-video Recommendation by Controlling Position Bias}
\toctitle{Improving Micro-video Recommendation by Controlling Position Bias}
% If the paper title is too long for the running head, you can set
% an abbreviated paper title here
%
\author{Yisong Yu\inst{1,2}
\and
Beihong Jin\inst{1,2(\textrm{\Letter})}
\and
Jiageng Song\inst{1,2}
\and
Beibei Li\inst{1,2}
\and
Yiyuan Zheng\inst{1,2}
\and
Wei Zhuo\inst{3}
}
% %
\authorrunning{Y. Yu et al.}
\tocauthor{Yisong Yu, Beihong Jin, Jiageng Song, Beibei Li, Yiyuan Zheng, Wei Zhuo}
% % First names are abbreviated in the running head.
% % If there are more than two authors, 'et al.' is used.
% %
\institute{State Key Laboratory of Computer Science, Institute of Software, Chinese Academy of Sciences, Beijing, China\\ \email{Beihong@iscas.ac.cn}\\ 
\and University of Chinese Academy of Sciences, Beijing, China
\and MX Media Co., Ltd., Singapore}
% \institute{Princeton University, Princeton NJ 08544, USA \and
% Springer Heidelberg, Tiergartenstr. 17, 69121 Heidelberg, Germany
% \email{lncs@springer.com}\\
% \url{http://www.springer.com/gp/computer-science/lncs} \and
% ABC Institute, Rupert-Karls-University Heidelberg, Heidelberg, Germany\\
% \email{\{abc,lncs\}@uni-heidelberg.de}}
%
\maketitle              % typeset the header of the contribution
\begin{abstract}
As the micro-video apps become popular, the numbers of micro-videos and users increase rapidly, which highlights the importance of micro-video recommendation. Although the micro-video recommendation can be naturally treated as the sequential recommendation, the previous sequential recommendation models do not fully consider the characteristics of micro-video apps, and in their inductive biases, the role of positions is not in accord with the reality in the micro-video scenario. Therefore, in the paper, we present a model named PDMRec (Position Decoupled Micro-video Recommendation). PDMRec applies separate self-attention modules to model micro-video information and the positional information and then aggregate them together, avoid the noisy correlations between micro-video semantics and positional information being encoded into the sequence embeddings. Moreover, PDMRec proposes contrastive learning strategies which closely match with the characteristics of the micro-video scenario, thus reducing the interference from micro-video positions in sequences. We conduct the extensive experiments on two real-world datasets. The experimental results shows that PDMRec outperforms existing multiple state-of-the-art models and achieves significant performance improvements.

\end{abstract}
\section{Introduction}
Micro-video streaming platforms are a kind of newly-emerging applications to provide entertainment services, where users can not only produce micro-videos and upload them to the platforms but also watch micro-videos in their spare time or fragmentary time by the apps on their smart phones. With micro-video apps, such as TikTok, Kwai, etc., increasingly popular, the numbers of micro-videos and users show a rapid growing trend.

In general, a micro-video app displays a single video in full-screen mode at a time and automatically plays it in a repetitive way. A user slides his finger on the screen to bring the next micro-video which is what the app recommends. If the micro-videos exposed to a user do not fall within the scope of his/her interests, then the user might give up watching and leave the app. Therefore, the recommendation, whose results will greatly affect the user experiences, becomes the key part in a micro-video streaming platform.

We note that micro-videos are large in quantity and short in time, usually in tens of seconds. Most of micro-videos are lack of the side information (such as genre, director, actors). By a micro-video app, a user can interact with a micro-video in several different ways, such as playing or replaying a micro-video, sharing or liking a micro-video, etc. These interactions can be divided into explicit feed-back (e.g. share or like) and implicit feedback (e.g. play or replay), where the former is few in number and the latter is large. In particular, while examining how the interaction sequences are generated, two phenomena need to be taken notice of. Firstly, the playing order of micro-videos is designated by the platform, instead of a result that a user proactively chooses. If the platform recommends another micro-video sequence, then a user has to browse or watch the micro-videos in that sequence in turn, leaving the totally different watching records. Secondly, the behaviors that a user browses micro-videos are driven by his/her interests, usually having no specific purposes, in comparison with the shopping in e-commerce scenarios. As a result, adjacent micro-videos in an interaction sequence have no strong inherent connections.

Existing micro-video recommendation models \cite{THACIL2018,ALPINE2019,MMGCN-MM2019,UVCAN-WWW2019,MTIN-MM2020} exploit multi-modal information including visual, acoustic, and textual features for recommending micro-videos. However, the multi-modal information of a micro-video is not always available, which renders these models impossible to effectively work. Even if having multi-modal information of each micro-video, we hold the view that these models are more appropriate for the micro-video ranking rather than the matching, where the matching and ranking are two standard successive phases in large-scale recommender systems. The reason behind our view is that the acquisition of multi-modal features requires a large amount of calculation. During the matching, the millions or even billions of micro-videos need to be processed, thus it is not practicable to recall micro-videos using these models in terms of time cost. Alternatively, sequential recommendation models \cite{GRU4Rec,GRU4RecDLRS,SRGNN2019,STAMP2018,SASRec2018,BERT4Rec2019,LightSANs2021} can be used for recalling micro-videos. However, so far, the characteristics inherent in interactions between a user and micro-videos have not been taken fully advantage of. Moreover, in most existing models \cite{GRU4Rec,STAMP2018,SASRec2018}, items in a user-item interaction sequence are believed to have unidirectional associations, that is, an item is dependent on one or more previous items in an interaction. Unfortunately, such an inductive bias is not matched exactly with what happens in the micro-video scenario.

In this paper, we propose a recommendation model named PDMRec (Position Decoupled Micro-video Recommendation) to recall micro-videos. We improve micro-video recommendations from two aspects. Firstly, we enhance the representation of the sequence embeddings by postponing the fusion of micro-video embedding and positional embedding so as to capture the nature of the order between micro-videos. Secondly, we argue that in the micro-video scenario, an original interaction sequence and its reordered sequence are semantically equivalent in the space of user interests, and further two reordered sequences from a same sequence are semantically equivalent. Thus, we can adopt the contrastive learning to reduce the interference from micro-video positions.

To summarize, the main contributions are as follows.
\littleshrinker
\begin{itemize}
\item[$\bullet$] We take a divide-and-merge policy to generate sequence embeddings, i.e., employ different multi-head attention to model micro-videos and their positional information in the sequences, respectively and then aggregate them into sequence embeddings, which can reflect the actual role of micro-video positions.
\item[$\bullet$] We construct semantically equivalent sequences by reordering operations for a given interaction sequence and present the reordering sequence loss for two newly-generated sequences, which can eliminate the implicit bias that the positional information brings out.
\item[$\bullet$] We conduct extensive experiments on two real-world datasets. The experimental results show PDMRec outperforms the other six models, i.e., GRU4Rec, STAMP, SASRec, BERT4Rec, CL4SRec and DuoRec, in terms of Recall and NDCG.
\end{itemize}

The rest of the paper is organized as follows. Section 2 introduces the related work. Section 3 describes the PDMRec model in detail. Section 4 gives the experimental evaluation. Finally, the paper is concluded in Section 5.

\section{Related Work}

Our work is related to the research under three non-orthogonal topics: micro-video recommendation, sequential recommendation and contrastive learning.

\littleshrinker
\subsubsection{Micro-video Recommendation.} Micro-video recommendation has attracted much attention of many researchers \cite{THACIL2018,ALPINE2019,MMGCN-MM2019,UVCAN-WWW2019,MTIN-MM2020}. For example, Chen et al .\cite{THACIL2018} characterizes both short-term and long-term correlations implied in user behaviors, and profiles user interests at both coarse and fine granularities. Li et al .\cite{ALPINE2019} present a temporal graph-based LSTM model to route micro-videos. Wei et al. \cite{MMGCN-MM2019} design a Multi-Modal Graph Convolution Network (MMGCN) which can yield modal-specific representations of users and micro-videos. Liu et al .\cite{UVCAN-WWW2019} propose the User-Video Co-Attention Network (UVCAN) which learns multi-modal information from both users and micro-videos using an attention mechanism. Jiang et al. \cite{MTIN-MM2020} propose a multi-scale time-aware user interest modeling framework, which learns user interests from fine-grained interest groups. These models, as we point out, utilize the multi-modal information of micro-videos but ignore the characteristics of interactions in micro-video apps.

\littleshrinker
\subsubsection{Sequential Recommendation.} Sequential recommendation is to model user interaction sequences to predict the next item the user will interact with, where an item can be a micro-video. A natural idea of modeling sequences is to employ RNNs. For example, GRU4Rec\cite{GRU4Rec} and its improved version \cite{GRU4RecDLRS} adopt multiple GRUs to predict the next item the user is most likely to interact with. Subsequently, various Graph Neural Networks (GNNs), such as SR-GNN \cite{SRGNN2019}, are applied to the sequential recommendation. However, comparing to the goods in e-commerce scenarios, micro-videos are akin to disposable products and might be forgotten by users quickly, which mismatches with the advantages of GNNs. The alternative way to achieve the sequential recommendation is to design attention mechanisms. For example, STAMP \cite{STAMP2018} provides an attention network to calculate the coefficient for each item in a sequence, and then generate general user interest and current user interest. Kang et al. apply Transformer \cite{Transformer2017} in NLP to the sequential recommendation and present SASRec \cite{SASRec2018}. SASRec adopts a self-attention mechanism to calculate the coefficients of items and presents positional embeddings to indicate the positions of items. Along with the advent of BERT \cite{BERT2019}, Sun et al. present BERT4Rec \cite{BERT4Rec2019}, which trains the bidirectional model to model sequential data using the cloze task. Fan et al. \cite{LightSANs2021} improve the self-attention module, which scales linearly w.r.t. the user’s historical sequence length in terms of time and space, and make the  model more resilient to over-parameterization.

% For example, SR-GNN \cite{SRGNN2019} builds item graphs over interaction sequences, and then learns embeddings of items on the graph and further embeddings of sequences. GNNs have the strength in modeling the different relationships between users and items. 

\littleshrinker
\subsubsection{Contrastive Learning.} More recently, contrastive learning has attracted a great deal of attention. It augments data to discover the implicit supervision signals in the input data and maximize the agreement between differently augmented views of the same data in a certain latent space. After the contrastive learning achieves first success in computer vision, it has been introduced to recommender systems \cite{CL4SRec,CLRec2021,CL-MM21,DuoRec2022,Contrastive-Learning-for-Denoising-Next-Basket-Recommendation}. For example, 
CL4SRec \cite{CL4SRec} proposes three methods to generate new sequences form raw data, and then  utilize them to improve the base model. DuoRec \cite{DuoRec2022} utilizes contrastive learning to resolve the representation degeneration, improving the recommendation accuracy.
Moreover, the contrastive learning also has been applied to reduce bias \cite{CLRec2021} and decrease noise \cite{Contrastive-Learning-for-Denoising-Next-Basket-Recommendation}, and alleviate the cold start problem \cite{CL-MM21}.

Compared to existing work, our model is tightly bound to the micro-video scenario. It generates embeddings of sequences in which positional information among micro-videos are encoded separately. It utilizes the scenario characteristic i.e., position independence to augment interaction sequences and narrow their semantic gap.

\section{Our Model}
%In this section, we first give the formulation of the problem to be solved. After outlining our model via the model architecture, we describe how we build components in the model and how we enable them to work together. We also analyze the space and time complexity and further discuss the features of our model design.

\littleshrinker
\subsection{Overview}
For a micro-video scenario, let $\mathcal{U}$ and $\mathcal{V}$ denote a user set and a micro-video set, respectively. For a user $u \in \mathcal{U}$, his/her positive interactions refer to the explicit/implicit feedback that meets some constraint conditions. The examples of positive interactions are given in Section $4.1$. We choose positive interactions from his/her watching records and order them by the interaction time, thus form a positive interaction sequence, denoted by $s_u=[v_1,v_2,...,v_{|s_u|}]$. 

\begin{figure}[tb]
\includegraphics[width=\textwidth]{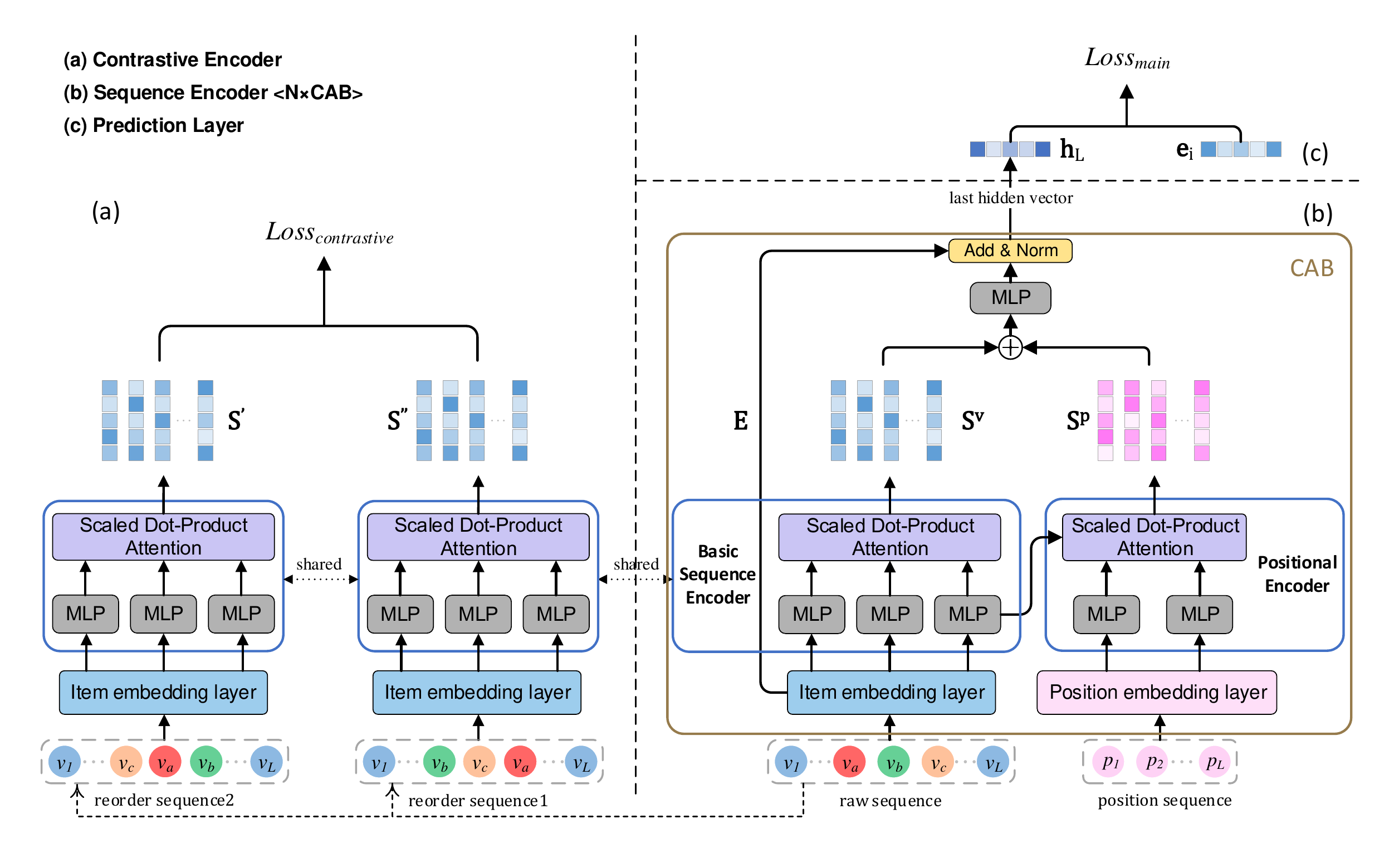}
\littleshrinker
\caption{Architecture of PDMRec.}
\label{fig1}
\littleshrinker
\end{figure}

Our goal is to predict the next micro-video that the user $u$ is most likely to be satisfied with, given the positive interaction sequence $s_u$ of the user $u$, where the criterion of user satisfaction is that his/her interaction with the micro-video belongs to a positive interaction.

Our model PDMRec exploits positive interaction sequences of users for recalling micro-videos, putting aside multi-modal information of micro-videos. For convenience of processing, we reconstruct each positive interaction sequence to be a sequence with fixed length $L$. Specifically, if a sequence has the length greater than or equal to $L$, then we choose latest $L$ interactions, otherwise we pad 0 from the end of sequence to the length $L$.  Hereinafter, we also denote this fixed-length sequence as $s_u$. 

Given the set $\mathcal{S}$ of fixed-length interaction sequences of users in $\mathcal{U}$, PDMRec learns a $d$-dimensional real-valued embedding $\mathbf{e}_i$ for each of the item $i$ in $\mathcal{V}$ and generates the sequence embedding matrix $\mathbf{S} \in \mathbb{R}^{d \times |\mathcal{S}|}$ by the sequence encoder, and enhances item embeddings by the contrastive encoder. Then in the prediction layer, PDMRec generates $\hat{y} = \{\hat{y}_1, \hat{y}_2, ..., \hat{y}_{|\mathcal{V}|}\}$ for each user, where $\hat{y}_i$ denotes the score for the item  $v_i$ in $\mathcal{V}$. Finally, PDMRec is trained as a classifier, taking the micro-videos with top-k scores as recommendations for each user. Fig. ~\ref{fig1} gives the architecture of the model. 

%\littleshrinker
\subsection{Sequence Encoder}
For generating sequence embeddings, the existing self-attention based recommendation models \cite{SASRec2018,BERT4Rec2019} adopt the method similar to that in Transformer \cite{Transformer2017} to gather the information of items in a sequence and positional information among these items. However, we find that these models take the addition of item embeddings and positional embeddings as the input of self-attention, which leads to  same projection matrices to be applied to different relationships, i.e., item-item  and position-position relationship, and will bring the mixed and noisy correlations. This defect is inherited from Transformer \cite{ke2021rethinking-positions}. We think it should be avoided in the scenarios of applying Transformer. 

In the sequence encoder, we introduce two slightly different attention modules, i.e., a basic sequence encoder and a positional encoder, to generate the basic sequence embedding and the positional embedding, respectively. 

In a basic sequence encoder, we use a fixed-length sequence $[v_1, v_2, ..., v_L]$ as the input of item embedding layer and obtain an item embedding matrix $\mathbf{E} = [\mathbf{e}_1, \mathbf{e}_2, ..., \mathbf{e}_L ], \mathbf{E} \in \mathbb{R}^{d \times L}$, where $d$ is the dimension of item embeddings. 

To capture the semantic dependence information among micro-videos in the sequence, we borrow a typical multi-head self-attention from Transformer and apply it with a different way from existing self-attention based recommendation models \cite{SASRec2018,BERT2019}. 

For the basic sequence embedding, we adopt the following multi-head attention module. 
\begin{equation}
% \begin{split}
% \begin{aligned}
\mathbf{S}_{h}^{v}=\operatorname{softmax}\left(\frac{\mathbf{E}^{T} \mathbf{W}_{Q h}\left(\mathbf{E}^{T} \mathbf{W}_{K h}\right)^{T}}{\sqrt{d / h d}}\right) \mathbf{E}^{T} \mathbf{W}_{V h}
\label{eq1}
% \end{aligned}
% \end{split}
\end{equation}
\begin{equation}
% \begin{split}
% \begin{aligned}
\mathbf{S}^v = \operatorname{concat}(\mathbf{S}^v_1, ..., \mathbf{S}^v_h)
\label{eq2}
% \end{aligned}
% \end{split}
\end{equation}

\noindent where $h \in [1, hd]$, $hd$ is the number of attention heads, $\mathbf{W}_{Qh}$, $\mathbf{W}_{Kh}$, $\mathbf{W}_{Vh} \in \mathbb{R}^{d \times dh}$ are learnable parameters, and $dh=d/hd$. 

Further, in the positional encoder, we introduce another self-attention module to exclusively handle positional embeddings for extracting the relationship among the positions of micro-videos. 

We take the position sequence as input of the position embedding layer and obtain the positional embedding matrix $\mathbf{P} \in \mathbb{R}^{d \times L}$, and then perform the following calculation. 

\begin{equation}
% \begin{split}
% \begin{aligned}
\mathbf{S}_{h}^{p}=\operatorname{softmax}\left(\frac{\mathbf{P}^{T} \mathbf{W}_{Q h}^{p}\left(\mathbf{P}^{T} \mathbf{W}_{K h}^{p}\right)^{T}}{\sqrt{d / h d}}\right) \mathbf{E}^{T} \mathbf{W}_{V h}
\label{eq3}
% \end{aligned}
% \end{split}
\end{equation}

\begin{equation}
% \begin{split}
% \begin{aligned}
\mathbf{S}^p = \operatorname{concat}(\mathbf{S}^p_1, ..., \mathbf{S}^p_h)
\label{eq4}
% \end{aligned}
% \end{split}
\end{equation}

\noindent where $h \in [1, hd]$, $hd$ is the number of attention heads, $\mathbf{W}^p_{Qh}$, $\mathbf{W}^p_{Kh} \in \mathbb{R}^{d \times dh}$ are learnable parameters, and $dh=d/hd$. Note that the number of head in the positional encoder is set to the same as the one in the sequence encoder. $\mathbf{S}^p$ is the resulting positional embeddings.

Having $\mathbf{S}^v$ and $\mathbf{S}^p$ in hand, we perform the following aggregation strategy.

\begin{equation}
% \begin{split}
% \begin{aligned}
\mathbf{S} = \operatorname{LayerNorm}(\operatorname{dropout}(\operatorname{MLP}(\mathbf{S}^v + \mathbf{S}^p)) + \mathbf{E}))
\label{eq5}
% \end{aligned}
% \end{split}
\end{equation}

As shown in Eq.~\ref{eq5}, we apply MLP, dropout, skip connection and layer normalization to aggregate $\mathbf{S}^v$ and $\mathbf{S}^p$, thus obtaining the sequence embedding matrix $\mathbf{S}$.

The basic sequence encoder, the positional encoder and the aggregator compose a context-aware block $(CAB)$. 

In the sequence encoder, we stack $N$ context-aware blocks $(CABs)$. Let $\mathbf{S}^1$ be the output of the first $CAB$ (i.e., $CAB^{(1)}$), the output of  $n$-th $CAB$ i.e., $CAB^{(n)}$ will be $\mathbf{S}^n = CAB^{(n)} (\mathbf{S}^{(n-1)}, \mathbf{P})$, where $n \in \{1, 2, …, N\}$ and $\mathbf{S}_0=\mathbf{E}$. $\mathbf{S}^N$ can be regarded as a set of $L$ hidden vectors, that is, $\mathbf{S}^N = [\mathbf{h}_1^N, \mathbf{h}_2^N, ..., \mathbf{h}_L^N]$, the hidden vector $\mathbf{h}_L^N$  is taken as the representation of the user sequence.

%\littleshrinker
\subsection{Contrastive Encoder}
In the micro-video scenario, we think the relative locations among a group of micro-videos that a user interacts with are not of great importance. Following the calculation in the sequence encoder, unidirectional associations between items are encoded into item embeddings, which is not amenable to the micro-video scenario. To eliminate impact of unidirectional association between items on item embeddings and reflect the essence of the order between micro-videos, we build a contrastive encoder using contrastive learning, which consists of two basic sequence encoders, as illustrated on the left side of Fig.~\ref{fig1}.

We generate new sequences which are semantically equivalent to the real interaction sequences by a reordering operation. More concretely, for a given positive interaction sequence $s_u$ of the user $u$, i.e., $s_u=[v_1, v_2, ..., v_{|s_u|}]$, we randomly shuffle a continuous sub-sequence $[v_r,v_{r+1}, ..., v_{r+L_r-1}]$, which starts at $r$ with length $L_r = \lfloor \alpha * |s_u| \rfloor$, to $[\hat{v}_r, \hat{v}_{r+1}, ..., \hat{v}_{r + L_r - 1}]$, where $\alpha$ is the proportion of reordering. As a result, we get the reordered sequence $s_r = [v_1, v_2, …, \hat{v_r}, ..., \hat{v}_{r+L_r-1}, \\ ..., v_{|s_u|}]$.

After applying the reordering operation twice, we generate two new sequences $s_{r1}$ and $s_{r2}$ for $s_u$. The reordering operation does not increase or decrease the number of items in the sequence, i.e., $|s_{r1}|=|s_{r2}|=|s_u|$. Next, we feed these two reordered sequences to the basic sequence encoders, respectively and obtain corresponding sequence representations $\mathbf{S}^{'} = [\mathbf{h}_1^{N'}, {\mathbf{h}_2^{N'}}, ..., \mathbf{h}_{|s_{u}|}^{N'}]$ and $\mathbf{S}^{''} = [\mathbf{h}_1^{N''}, \mathbf{h}_2^{N''}, ..., \mathbf{h}_{|s_{u}|}^{N''}]$. Here, we only send micro-video sequence information into the basic sequence encoder, deliberately ignoring the positional information of the sequence. This is for contrasting two augmented sequences without any disturbance from the positional information. This enables us to disentangle the contrastive loss and positional information modeling, as a result, the parameter updating inducing by the contrastive loss and the updating of positional embeddings do not affect each other.

Then, by the concatenation operation, we obtain $\hat{\mathbf{h}}^{'} = \operatorname{concat}(\mathbf{h}_1^{N'}, \mathbf{h}_2^{N'}, ..., \mathbf{h}_{|s_{u}|}^{N'})$ and $\hat{\mathbf{h}}^{''} = \operatorname{concat}(\mathbf{h}_1^{N''}, \mathbf{h}_2^{N''}, ..., \mathbf{h}_{|s_{u}|}^{N''})$. Finally, in order to minimize the gap between the representations of two sequences derived from the same original interaction sequence, we define the contrastive loss function in Eq.~\ref{eq6} as the reordering sequence loss, where we adopt the dot-product operation to calculate the similarity of two embeddings, i.e., for two embeddings $\mathbf{a}$, and $\mathbf{b}$, $\operatorname{sim}(\mathbf{a},\mathbf{b}) = \mathbf{a}^{T}\mathbf{b}$.

\begin{equation}
\setlength\abovedisplayskip{0pt}
\setlength\belowdisplayskip{0pt}
% \begin{split}
% \begin{aligned}
\mathcal{L}_{\mathrm{cl}}\left(\hat{\mathbf{h}}^{\prime}, \hat{\mathbf{h}}^{\prime \prime}\right)=-\log \frac{\exp \left(\operatorname{sim}\left(\hat{\mathbf{h}}^{\prime}, \hat{\mathbf{h}}^{\prime \prime}\right)\right)}{\exp \left(\operatorname{sim}\left(\hat{\mathbf{h}}^{\prime}, \hat{\mathbf{h}}^{\prime \prime}\right)\right)+\sum_{s_{*} \in S^{-}} \exp \left(\operatorname{sim}\left(\hat{\mathbf{h}}^{\prime}, \hat{\mathbf{h}}_{*}\right)\right)}
\label{eq6}
% \end{aligned}
% \end{split}
\end{equation}

While defining the reordering sequence loss, we think that it is more reasonable to take $\hat{\mathbf{h}}^{'}$, $\hat{\mathbf{h}}^{''}$, the results of concatenating all hidden vectors of the corresponding sequence representation, as parameters, because what we want the reordering sequence loss to do is to measure the semantic difference between two sequences. We have tried to replace $\hat{\mathbf{h}}^{'}$, $\hat{\mathbf{h}}^{''}$ with $\mathbf{h}^{N'}_L$, $\mathbf{h}^{N''}_L$ i.e., the last hidden vectors, which is proved ineffective by experimental results. 

Moreover, we adopt the following negative sampling method. Assume that there are $M$ sequences in a batch. Since we apply the reordering operation to the original sequence (e.g., $s_u$) to generate two new sequences (e.g., $s_{r1}$, $s_{r2}$) , we have a list of sequences with the length of $2M$. We treat $(s_{r1}$, $s_{r2})$ as a positive pair and the other sequences in the same batch as negatives, the latter forms the set of negatives, denoted by $S^-$.  $\hat{\mathbf{h}_*}$ is the concatenation of all the hidden vectors in the representation of $s_*$. 

%\littleshrinker
\subsection{Prediction and Loss Function}
We calculate the score for each candidate micro-video $v_i $ by conducting the dot product of sequence embedding $\mathbf{h}_L^n$ and embedding $\mathbf{e}_i$ of $v_i$, as shown in Eq.~\ref{eq7}.

\begin{equation}
% \begin{split}
% \begin{aligned}
\mathrm{Score}(v_i|\{v_1, …, v_L\}) = \operatorname{sim}(\mathbf{h}_L^N, \mathbf{e}_i)
\label{eq7}
% \end{aligned}
% \end{split}
\end{equation}

We adopt the negative log likelihood with full softmax as the main loss function. It can be written as follows.

\begin{equation}
% \begin{split}
% \begin{aligned}
\mathcal{L}_{\text {main }}=-\log \frac{\exp \left(\operatorname{sim}\left(\mathbf{h}_{L}^{N}, \mathbf{e}_{g}\right)\right)}{\sum_{i=1}^{|\mathcal{V}|} \exp \left(\operatorname{sim}\left(\mathbf{h}_{L}^{N}, \mathbf{e}_{i}\right)\right)}
\label{eq8}
% \end{aligned}
% \end{split}
\end{equation}

\noindent where $\mathbf{e}_g$ is the embedding of the ground-truth micro-video. Besides, we have the reordering sequence loss as an auxiliary loss. Finally, the total loss function is shown in Eq.~\ref{eq9}, where coefficient $\lambda$ is a hyperparameter.

\begin{equation}
% \begin{split}
% \begin{aligned}
\mathcal{L}_{total} = \mathcal{L}_{main} + \lambda \mathcal{L}_{cl}
\label{eq9}
% \end{aligned}
% \end{split}
\end{equation}

\littleshrinker
\subsection{Complexity Analysis}

\subsubsection{Space Complexity.}In PDMRec, the learnable parameters are mainly from the embeddings of micro-videos, i.e., \{$\mathbf{e}_i,i \in [1,|\mathcal{V}|]\}$, the positional embedding matrix $\mathbf{P} \in \mathbb{R}^{d \times L}$, the parameters in multi-head self-attention, i.e., $\mathbf{W}_{Qh}$, $\mathbf{W}_{Kh}$, $\mathbf{W}_{Vh}$,$\mathbf{W}^p_{Qh}$, $\mathbf{W}^p_{Kh} \in \mathbb{R}^{d \times dh}$, and the parameters in multiple MLPs, whose number are $|\mathcal{V}|\times d$, $Ld$, $d^2/hd$ and $d^2$, respectively. Therefore, the total space complexity of PDMRec is $O\left(|\mathcal{V}|d+Ld+d^2 \right)$.

For example, if training the model using WeChat dataset, the number of the learnable parameters is 2264144, where the number of item embedding is 2129472, which illustrates that item embeddings account for the bulk of all the parameters.

\subsubsection{Time Complexity.} The computation amount of PDMRec is mainly concentrated on the self-attention module and the MLPs, whose time complexity is $O\left( L^2d\right)$ and $O\left(Ld^2\right)$, respectively. Thus, the total time complexity is $O\left( L^2d+Ld^2\right)$.

A favorite property of PDMRec is that the computation in each self-attention module is fully parallelizable, which is suitable to GPU acceleration. Finally, it should be noted that our model does not increase the computational cost, compared to the state-of-the-art model such as DuoRec. 

%\littleshrinker
\subsection{Discussion}
\subsubsection{Avoiding Noisy Correlations.} Our model can avoid noisy correlations being involved into sequence embeddings. 
For Eq.~\ref{eq5}, we know sequence embeddings comes from the aggregation of $\mathbf{S}^v+\mathbf{S}^p$. Now, we expand $\mathbf{S}^v+\mathbf{S}^p$ in Eq.~\ref{eq10}. 

\begin{equation}
\begin{split}
\mathbf{S}^v+\mathbf{S}^p
&=\operatorname{concat}(\mathbf{S}_1^v,...,\mathbf{S}_h^v)+\operatorname{concat}(\mathbf{S}_1^p,...,\mathbf{S}_h^p)\\
&=\operatorname{concat}(\mathbf{S}_1^v+\mathbf{S}_1^p,...,\mathbf{S}_h^v+\mathbf{S}_h^p)
\label{eq10}
\end{split}
\end{equation}

We continue to expand Eq.~\ref{eq10} and obtain Eq.~\ref{eq11} as follows. For $j\in[1,h]$,
\begin{equation}
\begin{split}
\mathbf{S}^v_j+\mathbf{S}^p_j 
=&\operatorname{softmax}\left(\frac{\mathbf{E}^T\mathbf{W}_{Qj} \cdot (\mathbf{E}^T\mathbf{W}_{Kj})^{T}}{\sqrt{d /hd}}\right)\mathbf{E}^T\mathbf{W}_{Vj}\ + \\
&\operatorname{softmax}\left(\frac{\mathbf{P}^T\mathbf{W}^p_{Qj} \cdot (\mathbf{P^T}\mathbf{W}^p_{Kj})^{T}}{\sqrt{d / hd}}\right)\mathbf{E}^T\mathbf{W}_{Vj}\quad
\label{eq11}
\end{split}
\end{equation}

As shown in Eq.~\ref{eq11} , we apply linear projection matrices $\mathbf{W}_{Qj}$,$\mathbf{W}_{Kj}$ on item embedding matrix $\mathbf{E}$, apply linear projection matrices $\mathbf{W}^p_{Qj}$ and $\mathbf{W}^p_{Kj}$ on positional embedding matrix $\mathbf{P}$, and calculate the attention scores of $\mathbf{E}$ and $\mathbf{P}$ in different latent spaces. These projection matrices are different from each other and trained independently. Such the method adds no noisy correlations into sequence embeddings, reflecting the actual role of micro-video positions.

\subsubsection{Augmenting Equivalent Sequences.} For data augmentation method, besides the reordering operation, masking and cropping operations can also be applied on the sequences. Given a sequence $s_u$, the masking operation refers to randomly masking a proportion $\gamma$ of items and the cropping operation is to randomly delete a continuous sub-sequence of length $\eta * |s_u|$.
However, in the micro-video scenario, applying masking or cropping operations cannot guarantee generating semantically-equivalent sequences. Therefore, we do not choose these two operations. Experimental results in Section $4.4$ also illustrate that augmenting sequences by masking or cropping operations cannot achieve good recommendations.

\section{Experimental Evaluation}
%In this section, we conduct extensive experiments to answer the following three research questions:
% \textbf{RQ1}: Does PDMRec outperform existing recommendation models, whether classic models or state-of-the-art models?
% \textbf{RQ2}: What is the influence of various components in PDMRec?
% \textbf{RQ3}: How does the design of the contrastive learning strategies i.e., design of augmentation methods and contrastive loss, impact on the recommendation performance?

%\littleshrinker
\subsection{Experimental Setup}
\textbf{Datasets:} We adopt the following two datasets for experiments.
\begin{itemize}
\item[$\bullet$] \textbf{WeChat-Channels}: a dataset released by WeChat Big Data Challenge $2021$\footnote{https://algo.weixin.qq.com/problem-description}. It contains $14$-day interactions from anonymous users on WeChat Channels, a popular micro-video streaming platform in China.
\item[$\bullet$] \textbf{TikTok}: a dataset released by Short Video Understanding Challenge $2019$\footnote{https://www.biendata.xyz/competition/icmechallenge2019/} hosted by ByteDance, one of the largest companies in the world engaged in the micro-video streaming platform. It contains more than $275$ million user interactions with TikTok app.
\end{itemize}

Both datasets include engagement interactions such as watching and satisfaction interactions such as like and favorite. However, not every interaction between a user and a micro-video in the dataset reflect that the user likes the micro-video, so we set some constraints on interactions and choose those satisfying these constraints as positive interactions. More concretely, according to the characteristics of different micro-video streaming platforms (e.g. the average micro-video duration in WeChat and TikTok is 34.4s and 10.5s, respectively), we use different criteria to obtain positive interactions on TikTok and WeChat. 
From WeChat-Channels, we choose all the satisfaction interactions (such as likes and comments) and interactions whose loop times is greater than $1.1$ or watching time is greater than $45$s to form WeChat dataset. From TikTok, we randomly sample 10\% of users and choose their interactions whose loop times is greater than $1.0$, forming TikTok1. Also from TikTok, we choose all the interactions that users like micro-videos to form TikTok2 dataset. Next, we remove users and micro-videos with fewer than five  interactions, respectively so as to guarantee that each user/micro-video has enough interactions. Then, for each user, we sort his/her historical micro-videos by the interacted timestamp to obtain his/her interaction sequence. These sequences compose the datasets used in experiments. The statistics of processed datasets are shown in Table~\ref{tab1}.

We adopt the leave-one-out strategy to divide the dataset into train/validation/ test sets. That is, for each user $u$, we split his/her historical sequence $s = [v_1^u, v_2^u, ..., v_{|s|}^u ]$ into three parts: $(1)$ the most recent interaction $v_{|s|}^u$ for testing, $(2)$ the second most recent interaction $v_{|s|-1}^u$ for validation, and $(3)$ all remaining interactions for training.

 \begin{table}[tb]

    \renewcommand{\arraystretch}{1.4}
	\scriptsize
	\centering
	\caption{Statistics of datasets.}
	\label{tab1}
	\begin{tabular}{lccccc}
	\toprule
	Dataset & \#Users & \#Micro-videos & \#Interactions & Density & Avg. length of sequence\\ 
	\midrule
	WeChat & 19998 & 77985 & 2464798 & 0.16\% & 123.3\\
	TikTok1 & 19960  & 55954 & 2434289 & 0.22\% & 121.9\\
	TikTok2 & 30473 & 35716 & 684216 & 0.06\% & 25.5\\
	\bottomrule
	\end{tabular}
    \littleshrinker
\end{table}

\noindent \textbf{Metrics.} We employ two common top-k metrics, Recall@K and NDCG@K, to evaluate recommendation performance. Recall@K calculates the proportion of test items in top-k items of prediction scoring, while NDCG@K is a position-aware metric which assigns larger weights on higher positions. Since we only have one test item for each user, Recall@K is equivalent to Hit@K. In this paper, the value K is set to $20$, $50$, $100$.

In order to reduce the time cost of metric calculation, lots of previous work \cite{GRU4Rec,SASRec2018,BERT2019} adopts sampled negative items to calculate metrics. However, this method may lead to inconsistent with exact metrics \cite{sample-invalid-KDD20}. Therefore, we compute metrics on the whole item set to evaluate the model performance. That is, for each user, we rank all the micro-videos he/she has not interacted with by their scores rather than only rank the sampled negative items. 

\noindent \textbf{Implementation Details.} We implement our model with PyTorch, initializing all parameters by the normal distribution with mean $0$ and standard deviation $0.02$. Embedding size is set to $64$. The number of heads (i.e., hd) is set to $2$., the number of $CAB$ (i.e., $N$) is set to $2$. The proportion rate $\alpha$ of reordering is set to $0.2$. Coefficient $\lambda$ in the loss function is set to $0.1$. Dropout rate is set to $0.5$. We use Adam as the optimizer with the learning rate of $0.001$. Batch size (i.e., $M$) is set to $512$. The sequence length $L$ is set to $100$ for WeChat and TikTok1 datasets, and $50$ for TikTok$2$ dataset. Our code is available publicly on GitHub\footnote{https://github.com/Ethan-Yys/PDMRec} for reproducibility.

We train the model by an early stopping technique, that is, when Recall$@50$ on the validation set has not been improved in $15$ consecutive epochs, we stop training the model. 

\subsection{Performance Comparison}
\subtleshrinker
We conduct comparative experiments, comparing our model with the following six models.
% to answer \textrm{RQ1}.
\begin{itemize}
\item[$\bullet$] GRU4Rec$^+$ \cite{GRU4RecDLRS}. It uses GRU modules to model user preferences by interaction sequences and is improved with prefix sub-sequences as data augmentation and a method to account for shifts in the input data distribution.
\item[$\bullet$] STAMP \cite{STAMP2018}. It is a short-term attention/memory priority model which aims to capture long-term user preferences from previous interactions and short-term user preferences from the last interaction in a sequence.
\item[$\bullet$] SASRec \cite{SASRec2018}. It firstly introduces the Transformer encoder to learn user representations. It models use preferences through a self-attention mechanism and achieves state-of-the-art performance at that time.
\item[$\bullet$] BERT4Rec\cite{BERT4Rec2019}. It learns via the BERT structure, that is, employs the deep bidirectional self-attention to model user behavior sequences. 
\item[$\bullet$] CL4SRec \cite{CL4SRec}. It firstly uses contrastive learning to enhance user representations in sequential recommendation. Specifically, item cropping, masking and reordering are applied to an original sequence to generate sequences for calculating the contrastive loss.

\item[$\bullet$] DuoRec \cite{DuoRec2022}. It is also a method based on contrastive learning. It uses both supervised and unsupervised methods to generate sequences for calculating contrastive loss.
\end{itemize}

Among these competitors, GRU4Rec$^+$, STAMP, SASRec and BERT4Rec are from a popular open-source recommendation framework RecBole\footnote{https://github.com/RUCAIBox/RecBole}. CL4SRec and DuoRec are from the implementation of Zhao et al.\footnote{https://github.com/RuihongQiu/DuoRec} To be fair, in all the models, we set the dimension of item embedding to $64$ and batch size to $512$. Particularly, we adopt the $2$-layer encoder for the models which apply self-attention.

\begin{table}[htb]
\scriptsize
\renewcommand{\arraystretch}{1.4}
\centering
\caption{Recommendation performance on three datasets.} 
\label{tab2}
\leftline{\textbf{WeChat}}
\begin{tabular}{l|l|cccccccc}
\toprule
\multicolumn{2}{c|}{Metrics} & GRU4Rec$^+$ & STAMP & SASRec & BERT4Rec & CL4SRec & DuoRec & PDMRec & Improv.(\%) \\ 
\midrule
&\textbf{@20} & 0.1093 & 0.0888 & 0.1069 & 0.0892 & 0.1035 & \underline{0.1108} & \textbf{0.1157} & 4.42\% \\
\textbf{Recall}&\textbf{@50} & \underline{0.2125} & 0.1727 & 0.2095 & 0.1762 & 0.2057 & 0.2169 & \textbf{0.2224} & 2.54\% \\
&\textbf{@100} & \underline{0.3270} & 0.2647 & 0.3224 & 0.2744 & 0.3161 & 0.3263 & \textbf{0.3423} & 4.90\% \\
\midrule
&\textbf{@20} & 0.0407 & 0.0339 & 0.0388 & 0.0322 & 0.0386 & \underline{0.0410} & \textbf{0.0422} & 2.93\% \\
\textbf{NDCG}&\textbf{@50} & 0.0610 & 0.0504 & 0.0591 & 0.0494 & 0.0587 & \underline{0.0619} & \textbf{0.0632} & 2.10\% \\
&\textbf{@100} & 0.0795 & 0.0652 & 0.0774 & 0.0654 & 0.0766 & \underline{0.0796} & \textbf{0.0826} & 3.77\% \\
\bottomrule
\end{tabular}

\flushleft{\textbf{TikTok1}}
%\leftline{\textbf{TikTok1}}
\begin{tabular}{l|l|cccccccc}
\toprule
\multicolumn{2}{c|}{Metrics} & GRU4Rec$^+$ & STAMP & SASRec & BERT4Rec & CL4SRec & DuoRec & PDMRec & Improv.(\%) \\ 
\midrule
&\textbf{@20} &0.1003 & 0.0815 & 0.0112 & 0.0842 & 0.1066 & \underline{0.1139} & \textbf{0.1190} & 4.48\% \\
\textbf{Recall}&\textbf{@50} & 0.1915 & 0.1503 & 0.2028 & 0.1609 & 0.2005 & \underline{0.2071} & \textbf{0.2200} & 6.23\% \\
&\textbf{@100} & 0.2991 & 0.2308 & 0.3087 & 0.2525 & 0.3062 & \underline{0.3114} & \textbf{0.3281} & 5.36\% \\
\midrule
&\textbf{@20} & 0.0395 & 0.0342 & 0.0462 & 0.0319 & 0.0424 & \underline{0.0465} & \textbf{0.0481} & 3.44\% \\
\textbf{NDCG}&\textbf{@50} & 0.0574 & 0.0477 & 0.0640 & 0.0470 & 0.0609 & \underline{0.0648} & \textbf{0.0687} & 6.02\% \\
&\textbf{@100} & 0.0748 & 0.0607 & 0.0812 & 0.0618 & 0.0780 & \underline{0.0817} & \textbf{0.0861} & 5.39\% \\
\bottomrule
\end{tabular}
% \tinyshrinker

\flushleft{\textbf{TikTok2}}
%\leftline{\textbf{TikTok2}}
\begin{tabular}{l|l|cccccccc}
\toprule
\multicolumn{2}{c|}{Metrics} & GRU4Rec$^+$ & STAMP & SASRec & BERT4Rec & CL4SRec & DuoRec & PDMRec & Improv.(\%) \\ 
\midrule
&\textbf{@20} & 0.0514 & 0.0422 & 0.0777 & 0.0623 & 0.0727 & \underline{0.0808} & \textbf{0.0830} & 2.70\% \\
\textbf{Recall}&\textbf{@50} & 0.0979 & 0.0795 & 0.1240 & 0.1045 & 0.1177 & \underline{0.1266} & \textbf{0.1332} & 5.20\% \\
&\textbf{@100} & 0.1547 & 0.1213 & 0.1752 & 0.1548 & 0.1698 & \underline{0.1781} & \textbf{0.1894} & 6.34\% \\
\midrule
&\textbf{@20} & 0.0205 & 0.0175 & 0.0475 & 0.0314 & 0.0432 & \underline{0.0479} & \textbf{0.0491} & 3.55\% \\
\textbf{NDCG}&\textbf{@50} & 0.0297 & 0.0248 & 0.0566 & 0.0398 & 0.0520 & \underline{0.0569} & \textbf{0.0590} & 3.69\% \\
&\textbf{@100} & 0.0388 & 0.0316 & 0.0645 & 0.0479 & 0.0604 & \underline{0.0653} & \textbf{0.0681} & 4.29\% \\
\bottomrule
\end{tabular}
\end{table}

Experimental results on WeChat, TikTok$1$ and TikTok$2$ are listed in Table~\ref{tab2}, where the number in a bold type is the best performance in each row and the underlined number is the second best in each row.

From the results, we have the following observations.
\begin{itemize}
\item[$\bullet$] The attention-based sequential recommendation models are inferior to ones which integrate additional contrastive learning module in terms of both Recall and NDCG. It makes us believe that augmenting data more targeted for the specific scenario and optimizing in different latent spaces do improve performance. 
\item[$\bullet$] The classic baseline i.e., GRU4Rec$^+$ performs not bad, which is beyond our expectation. As shown in Table~\ref{tab2}, the worse one in performance on three datasets is STAMP instead of GRU4Rec$^+$. In particular, on datasets WeChat and TikTok$1$ which contain not only explicit feedback but also implicit feedback, GRU4Rec$^+$ is even superior to several attention-based models such as SASRec. It is supposed that GRU4Rec$^+$ is more appropriate to modeling long sequences. 
\item[$\bullet$] Most importantly, PDMRec surpasses all the competitors in all metrics on the three datasets. For example, on the TikTok$1$ dataset, PDMRec outperforms the second best model, i.e., DuoRec, about $6.23\%$ on Recall$@50$ and $6.02\%$ on NDCG$@50$. We think the good performance of PDMRec stems from two steps of position decoupling: learning positional embedding independently and optimizing representations of reordered sequences for position-independence semantic conformity.
\end{itemize}

\subsection{Ablation Study}
\begin{figure}
\includegraphics[width=\textwidth]{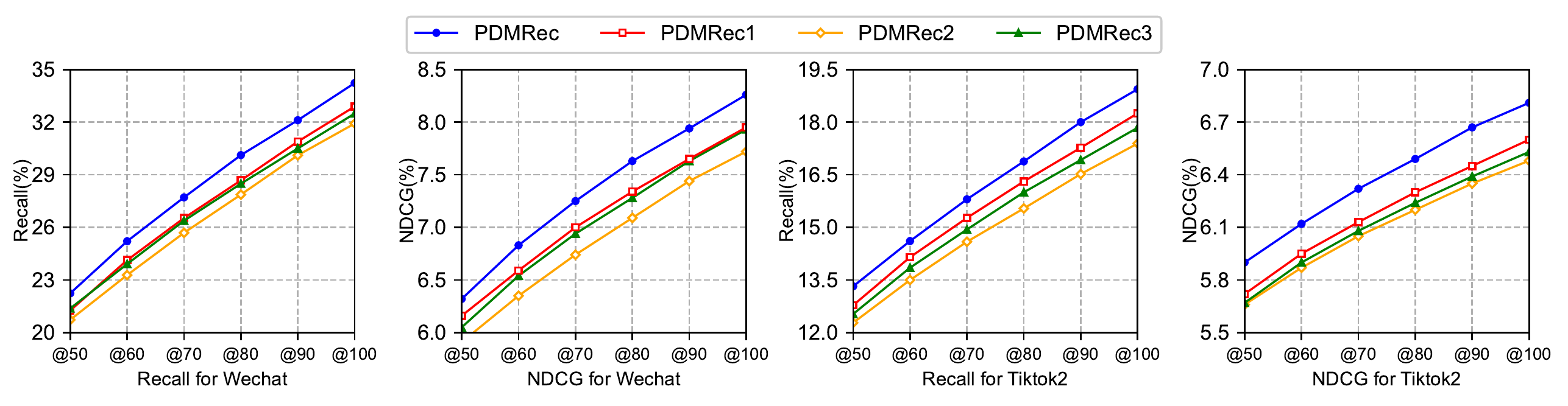}
\caption{Ablation study on two datasets} 
\label{fig2}
\end{figure}

We conduct the ablation study on our model to observe the effectiveness of different components.
%and answer \textrm{RQ2}.
 We compare our model with three variants, i.e., PDMRec$1$, PDMRec$2$ and PDMRec$3$. PDMRec$1$ and PDMRec$2$ are PDMRec models which remove the contrastive encoder and the positional encoder, respectively. PDMRec$3$ is the PDMRec model which removes the positional encoder but adopts the addition of item embedding and positional embedding as the input of the model. The results are shown in Fig.~\ref{fig2}.

From Fig.~\ref{fig2}, we find the variant PDMRec$1$ suffers severe declines in performance, which illustrates that optimizing the reordering sequence loss has a favorable effect on improving the performance. Moreover, PDMRec$2$ shows the worst performance while comparing with PDMRec$1$ and other variants. It proves that lacking position information will lead to sharply declining in performance. Further, we compare the performance of PDMRec and PDMRec$3$, where they provide the different ways of fusing position information into sequence embeddings. Obviously, PDMRec has better performance, which shows that encoding positional embeddings separately is necessary.

\subsection{Impact of Contrastive Learning Strategies}
We conduct experiments to evaluate the rationality of contrastive learning strategies adopted in this paper.
%and answer \textrm{RQ3}.
The contrastive learning strategies involve data augmentation and contrastive learning loss calculation. 

\begin{table}[b]
\scriptsize
\renewcommand{\arraystretch}{1.4}
\centering
\caption{Recommendation performance of different contrastive learning strategies.} 
\label{tab6}
% \footnotesize
% \tinyshrinker
\begin{tabular}{l|l|cccccc}
\toprule
\multicolumn{2}{c|}{Metrics} & PDMRec & PDMRec4 & PDMRec5 & PDMRec6 & PDMRec7 & PDMRec8\\
\midrule
&\textbf{@20} & \textbf{0.1157} & 0.1118 & 0.1068 & 0.1103 & 0.1081 & 0.1099 \\
\textbf{Recall}&\textbf{@50} & \textbf{0.2224} & 0.2157 & 0.2144 & 0.2138 & 0.2134 & 0.2183 \\
&\textbf{@100} & \textbf{0.3423} & 0.3277 & 0.3311 & 0.3309 & 0.3297 & 0.3282 \\
\midrule
&\textbf{@20} & \textbf{0.0422} & 0.0416 & 0.0393 & 0.0400 & 0.0393 & 0.0405 \\
\textbf{NDCG}&\textbf{@50} & \textbf{0.0632} & 0.0620 & 0.0606 & 0.0604 & 0.0600 & 0.0619 \\
&\textbf{@100} & \textbf{0.0826} & 0.8010 & 0.0794 & 0.0794 & 0.0788 & 0.0796 \\
\bottomrule
\end{tabular}
% \tinyshrinker
\end{table}

By replacing the reordering operation in PDMRec with other operations in CL4SRec, we can obtain different variants: PDMRec$4$ using the masking operation, PDMRec5 using the cropping operation, and PDMRec6 using the masking, cropping and reordering operations. In experiments, we set $\gamma$ to $0.5$ and $\eta$ to $0.5$, the values which lead to the best performance of experiments in \cite{CL4SRec}. The performance of these variants is listed in the columns $3$-$5$ of Table~\ref{tab6}. From the results, we find that three variants fall far behind PDMRec in terms of Recall and NDCG. These results illustrate that using the reordering operation to generate new sequences is effective. 

For the contrastive learning loss calculation, in our model, we use $\hat{\mathbf{h}}^{'}$ and $\hat{\mathbf{h}}^{''}$, which are the output of the Scaled Dot-Product Attention in the basic sequence encoder, to calculate the loss. An alternative is to take last vector $\hat{\mathbf{h}}^{N'}_L$, $\hat{\mathbf{h}}^{N''}_L$ in ${\mathbf{S}}^{'}$ and ${\mathbf{S}}^{''}$ to calculate the loss, which is the way CL4SRec adopts. Another alternative is to take the output of Add$\&$Norm layer to calculate the loss, which is the way DuoRec adopts. The corresponding variants are denoted by PDMRec$7$ and PDMRec$8$. The performance of these variants is listed in the columns 6-7 of Table~\ref{tab6}.

We find PDMRec significantly outperforms PDMRec$7$. It illustrates using all the hidden vectors to calculate the loss is able to optimize augmented sequences more comprehensively. Moreover, while comparing to PDMRec, PDMRec$8$ has a relatively low performance, which shows that our design avoids positional errors and has a positive effect on performance.

\section{Conclusion}
In this paper, we examine the characteristics of the micro-video scenario, rethink the role of positions in interaction sequences, and then propose a micro-video recommendation model PDMRec. For improving the representations of sequences and avoiding the noise being added into, PDMRec encodes micro-videos and positions in sequences, respectively, computing the micro-video contextual correlations and positional correlations with different parameterizations. Further, PDMRec adopts the reordering operation to augment interaction sequences and presents a reordering sequence loss to remedy the negative impact brought by micro-video positions in sequences. Results of experiments on real-world datasets show that our PDMRec model is effective in terms of Recall and NDCG, and outperforms the state-of-the-art baselines.

\subsubsection{Acknowledgements} This work was supported by the National Natural Science Foundation of China under Grant No. 62072450 and the 2021 joint project with MX Media.

%
% ---- Bibliography ----
%
% BibTeX users should specify bibliography style 'splncs04'.
% References will then be sorted and formatted in the correct style.
%
% \bibliographystyle{splncs04}
% \bibliography{mybibliography}
%
\littleshrinker

\begin{comment}
{\small
	\bibliographystyle{splncs04}
	\bibliography{reference}
}
\end{comment}

\end{document}